\documentclass{elsart}
\usepackage{longtable}
\usepackage{booktabs}

\begin{document}
\newcommand{\xbeq}{\begin{eqnarray}} \newcommand{\xeeq}{\end{eqnarray}}

\runauthor{}
\begin{frontmatter}
  \title{Mie Scattering by Ensembles of Particles with Very Large Size Parameters}
  \author[Caltech,MPIA]{S.\ Wolf},
  \author[SPBU]{N.\ V.\ Voshchinnikov} 

  \address[Caltech]{California Institute of Technology, 1201 East California Blvd, 
	Mail Code 105-24, Pasadena CA 91125,
    	USA}
  \address[MPIA]{Max Planck Institute for Astronomy, K\"onigstuhl 17, 69117
	Heidelberg, Germany, swolf@mpia.de}
  \address[SPBU]{Sobolev Astronomical Institute, St.~Petersburg University,
    Universitetskii prosp. 28, 198504 Stary Peterhof-St.~Petersburg, Russia,
    nvv@astro.spbu.ru}
	
  \begin{abstract}
  We present a computer program for the simulation of Mie scattering in case of arbitrarily
  large size parameters. The elements of the scattering matrix, efficiency factors 
  as well as the corresponding cross sections, the albedo
  and the scattering asymmetry parameter are calculated.
  Single particles as well as particle ensembles consisting of
  several components and particle size distributions can be considered.

  \vspace*{4mm}

  \vspace*{-4mm}
  \end{abstract}
\end{frontmatter}

\newpage
\section{Program summary}

{\sl Title of program:} {\bf\em miex}

{\sl Catalogue identifier: } 

{\sl Program obtainable from:} CPC Program Library, Queen's University of Bel\-fast, N. Ireland

{\sl Computer for which the program is designed and others on which it has been tested:}

{\sl Computers: } 
Any machine running standard FORTRAN 90; 
{\bf\em miex} has been tested on 
an Intel Celeron processor (Redhat Linux 9.0, Intel Fortran Compiler 7.1),
an Intel XEON processor    (SuSE   Linux 9.0, Intel Fortran Compiler 8.0), and
a Sun-Blade-1000           (OS 8.5, Sun Workshop Compiler Fortran 90 2.0).

{\sl Installations:} standard

{\sl Operating systems or monitors under which the program has been tested:}
Redhat Linux 9.0, 
SuSE   Linux 9.0,
Sun OS 8.5

{\sl Programming language used:} Fortran 90

{\sl Memory required to execute with typical data:} 
1\,MByte - several 100\,MByte (see Sect.~\ref{testrun} for examples)

{\sl No. of bits in a word:} 8

{\sl No. of processors used:} 1

{\sl Has the code been vectorized or parallelized ?:} NO

{\sl No. of bytes in distributed program, including test data, etc.:}

{\sl Distributed format:}  ASCII

{\sl Keywords :}
Mie scattering, size parameter, particle ensemble, scattering matrix,
Stokes vector, efficiency factor, cross section, asymmetry
parameter, albedo

{\sl Nature of the physical problem:} 
Among a variety of applications, Mie scattering is of essential
importance for the continuum radiative transfer in cosmic dust configurations.
In this particular case, Mie theory describes the interaction
of electromagnetic radiation with spherical dust grains on the basis of their 
complex refractive index and size parameter.
Both, broad grain size distributions (radii $a$: nanometers -- millimeters)
and a very wide wavelength range ($\lambda \approx 10^{-10}-10^{-2}$\,m)
of the interacting radiation are  considered. Previous numerical solutions
to the Mie scattering problem are not appropriate to consider
size parameters $x=2\pi a/\lambda >10^4 - 10^5$.
In contrast to this, the presented code allows to consider arbitrary
size parameters. It will be useful not only for
applications in astrophysics
but also in other fields of science
(atmospheric and ocean optics, biophysics, etc.) and industry (particle sizing, ecology
control measurements, etc.).

{\sl Method of solution:} 
Calculations of Mie scattering coefficients
and efficiency factors as outlined by Voshchinnikov (2004), combined
with standard solutions of the scattering amplitude functions.
Single scattering by particle ensembles is calculated by proper averaging of the respective
parameters.

{\sl Restrictions on the complexity of the problem:} 
Single Scattering

{\sl Typical running time:}
Seconds to minutes.

{\sl Unusual features of the program:} 
None

{\sl References:}\\
Bohren C.F., Huffman D.R.,
Absorption and scattering of light by small particles.
John Wiley \& Sons, New York (1983)

Voshchinnikov N.V.:
``Optics of Cosmic Dust. I", {\sl Astrophysics and Space Physics Reviews} 12,  1 (2004)

\newpage
\section{Introduction}

The motivation for the development of the presented Mie scattering routine
was given by the goal to model the continuum radiative transfer in the circumstellar environment
of young stellar objects and X-ray halos where dust scattering and thermal
dust reemission dominate the continuum radiation field.
In the most approaches, dust grains are assumed to be homogeneous spheres.
Their optical properties are defined through Mie theory with following input parameters:
(a) the complex refractive index $m = n + i k$ of a material relative to the surrounding medium
and 
(b) the dimensionless size parameter
\begin{equation}\label{grpar}
  x = \frac{2\pi a}{\lambda},
\end{equation}
connecting the particle radius $a$ and
the wavelength of the incident electromagnetic radiation $\lambda$.
New findings suggest that the radii of the largest grains
in circumstellar disks around young stellar objects may reach several millimeters to 
centimeters in radius (see, e.g., Calvet et al.~2002). 
Since the embedded stars emit a considerable amount of energy
even in the ultraviolet wavelength range (effective temperature of a
typical T\,Tauri star:
$T_{\rm eff}$=4000\,K -- Gullbring et al.~1998), the resulting size parameter $x$ amounts up to $10^5 - 10^6$.
In more evolved circumstellar disks, so-called ``debris disks'', the size parameter $x$
may even exceed this value by several orders of magnitude.
Furthermore, interstellar dust grains located near the line of sight toward
distant X-ray sources will scatter some of its radiation and thus produce a diffuse halo around a point-like source.
Observations are performed at energies $E \sim 10$\,keV (or more see, e.g., Woo et al., 1994) that
correspond to wavelengths $\lambda \sim 1.24$\,\AA. In this domain, the size parameter is equal to
$x = 5067 (a/\mu{\rm m}) (E/{\rm keV})$ and exceeds $5\times 10^6$ if $a \ge 100\,\mu$m.

For the solution of the described radiative transfer problems, the efficiency factors
and scattering matrix are required. Under the assumption
of spherical, homogeneous particles,
these quantities are calculated from the Mie series solution.
The most widely applied numerical Mie code is that from Bohren \& Huffman (1983).
It is based on the improved  algorithm of Mie scattering of
Wiscombe~(1980, 1996) and is restricted to size parameters $\le 2\times 10^4$.
Therefore, this numerical solution is not appropriate for the applications described above
(see also Shah~1977, 1992 for Mie calculations in the size parameter regime $x \leq 10^5$).

The computer program presented in this article is based on the calculation of
the Mie  coefficients and efficiency factors for single particles
as described by Voshchinnikov (2004).
The calculations of the angular functions are based on the standard approach
(see Wiscombe~1980, Bohren \& Huffman 1983).
We successfully tested the direct implementation of
Legendre functions (Zhang \& Jin~1996) 
for the determination of the scattering amplitude functions
as well, but only the fastest,
i.e., the first approach is realized in the published version of the code.
To account for particle size distributions and/or ensembles of
different components,
a proper averaging of the single parameters (efficiency factors etc.)
is implemented as well.

\newpage
\section{Long Write-up}

\subsection{Mie scattering}

In the following, a brief overview of the scattering process and
the different quantities which can be
derived with {\bf\em miex} is given. For a more detailed description of
the Mie scattering solution we refer to the books of
van de Hulst (1957) and
Bohren \& Huffman (1983).

Let the radiation be described by the four-component Stokes vector $\hat{I}=(I,Q,U,V)^{\rm T}$.
The scattering of radiation by a spherical particle is described by
the scattering (M\"uller)  matrix $\hat{F}$ of special type
\begin{equation}\label{scmat}
\hat{I_1} \propto \hat{F} \hat{I_0},
\end{equation}
where $\hat{I_0}$ ($\hat{I_1}$) is the Stokes vector before (after) the scattering and
\begin{equation}\label{kugscmat}
  \hat{F}(\Theta) =
  \left(
    \begin{array}{cccc}
      F_{11}(\Theta) & F_{12}(\Theta) &  0              & 0\\
      F_{12}(\Theta) & F_{11}(\Theta) &  0              & 0\\
      0              & 0              &  F_{33}(\Theta) & F_{34}(\Theta)\\
      0              & 0              & -F_{34}(\Theta) & F_{33}(\Theta)
    \end{array}
  \right).
\end{equation}
Here, $\Theta$ is the angle between the direction of the incident and the scattered
radiation (scattering angle).
The elements of
the scattering matrix $F_{ik}$ can be derived from the complex
amplitude functions (the asterisk denotes the complex conjugation)
$S_{1}$, $S_{2}$:
\xbeq
\begin{array}{c}
F_{11}(\Theta)  =  \displaystyle \frac{1}{2}\left[ \left| S_2(\Theta) \right|^2   +  \left| S_1(\Theta) \right|^2\right],
\hspace*{0.5cm}
F_{12}(\Theta)  =  \displaystyle\frac{1}{2}\left[ \left| S_2(\Theta) \right|^2   -  \left| S_1(\Theta) \right|^2\right],\\
F_{33}(\Theta)  =  \displaystyle\frac{1}{2}\left[S^*_2(\Theta) S_1(\Theta)   + S_2(\Theta) S^*_1(\Theta)\right], \\ \\
\smallskip
F_{34}(\Theta)  =  \displaystyle\frac{i}{2}\left[S_1(\Theta)   S^*_2(\Theta) - S_2(\Theta) S^*_1(\Theta)\right].\\
\end{array}
\xeeq
The amplitude functions can be calculated as follows:
\begin{equation}\label{s1}
S_{\rm 1}(\Theta)  = \sum^{\infty}_{n=1} \frac{2n+1}{n(n+1)}\left[a_n\pi_n(\Theta)  + b_n\tau_n(\Theta)\right]
\end{equation}
and
\begin{equation}\label{s2}
S_{\rm 2}(\Theta) = \sum^{\infty}_{n=1} \frac{2n+1}{n(n+1)}\left[a_n\tau_n(\Theta) + b_n\pi_n(\Theta)\right].
\end{equation}
The complex scattering (Mie) coefficients $a_n$ and $b_n$
depend on the size parameter $x$ and the refractive index $m$ of the material.
The angular functions $\pi _{n}$ and $\tau_{n}$
depend on $\cos \Theta $ only and can be found from recurrence relations
(Wiscombe~1980; $n \geq 2$)
\begin{equation}\label{pi}
\pi_{n}(\Theta) = \frac{2n-1}{n-1}\cos \Theta \, \pi_{n-1}(\Theta) -
                  \frac{n}{n-1}\pi_{n-2}(\Theta)
\end{equation}
and
\begin{equation}\label{tau}
\tau_{n}(\Theta) = n \cos \Theta \pi_{n}(\Theta) -(n+1) \pi_{n-1}(\Theta).
\end{equation}
The initial values are
\begin{equation}\label{pi1}
\pi _{0}(\Theta) = 0, \qquad \pi _{1}(\Theta) = 1.
\end{equation}
The coefficients 
\begin{equation}
a_{n} = \frac{\psi'_n(mx)\psi_{n}(x) - m\psi_{n}(mx)\psi'_n(x)}
             {\psi'_n(mx)\zeta_{n}(x) - m\psi_{n}(mx)\zeta'_n(x)}
\end{equation}
and
\begin{equation}
b_{n} = \frac{m\psi'_n(mx)\psi_{n}(x) - \psi_{n}(mx)\psi_n'(x)}
              {m\psi'_n(mx)\zeta_{n}(x)-\psi_{n}(mx)\zeta_n'(x)}
\end{equation}
are usually transformed to a form convenient for calculations.
Following Deirmendjian~(1969) and Loskutov~(1971), we
replace the Riccati-Bessel functions $\psi_n(x)$ and $\zeta_n(x)$
and their first derivatives  $\psi'_n(x)$ and $\zeta'_n(x)$
by spherical Bessel functions of the first and second kind
$J_{n+1/2}(x)$ and $Y_{n+1/2}(x)$. With the aid of recursive
relations for these functions
(see, e.g., Abramowitz and Stegun~1964) we obtain:
\begin{equation}
\hspace*{-0.9cm} a_{n} = \displaystyle \frac{ \left[\frac{{\mathcal A}_n}{m} + \frac{n}{x} \right]
            J_{n+1/2}(x) - J_{n-1/2}(x)}
{\left[\frac{{\mathcal A}_n}{m} + \frac{n}{x} \right] J_{n+1/2}(x) - J_{n-1/2}(x) +
 i \left\{ \left[\frac{{\mathcal A}_n}{m} + \frac{n}{x} \right] Y_{n+1/2}(x) - Y_{n-1/2}(x)
 \right\}}
\end{equation}
and
\begin{equation}
\hspace*{-0.9cm}
b_{n} =
 \frac{\left[m{\mathcal A}_n + \frac{n}{x} \right]
            J_{n+1/2}(x) - J_{n-1/2}(x)}
{\left[ m{\mathcal A}_n + \frac{n}{x} \right] J_{n+1/2}(x) - J_{n-1/2}(x) +
 i \left\{ \left[m{\mathcal A}_n+\frac{n}{x} \right] Y_{n+1/2}(x) - Y_{n-1/2}(x)
 \right\} },
\end{equation}
where
\begin{equation}
{\mathcal A}_{n} \equiv  \frac{\psi'_n(mx)}{\psi _{n}(mx)} =
 - \frac{n}{mx} + \frac{J_{n-1/2}(mx)}{J_{n+1/2}(mx)}.
\end{equation}
The logarithmic derivative to the Bessel functions of a complex argument
${\mathcal A}_{n}$
is calculated via backward recursion using the relation
\begin{equation}
{\mathcal A}_n =
 - \frac{n}{mx} + \left(\frac{n}{mx}-{\mathcal A}_{n-1} \right)^{-1}.
\end{equation}
The starting number (variable ``{\tt num}'' in the program) for $x \leq 50000$
(${\mathcal A}_{n}=0$)
is chosen according to the recommendation of
Loskutov~(1971).
It  is smaller for large arguments than that given by Wiscombe~(1980).

According to Voshchinnikov (2004), the calculation of the Bessel functions
of a real argument
is based on the upward recursion for the functions of the second kind $Y_{n+1/2}(x)$,
which is known to be stable.
This is given by the relation
(Abramowitz and Stegun~1964; $n \geq 0$):
\begin{equation}
Y_{n+5/2}(x) = \frac{2n+1}{x}Y_{n+3/2}(x) - Y_{n+1/2}(x)
\label{yy}\end{equation}
with the initial values
\begin{equation}\label{y0}
Y_{1/2}(x)  = -\sqrt{\frac{2}{\pi x}} \cos x, \qquad
Y_{3/2}(x)  = \sqrt{\frac{2}{\pi x}} \left(- \frac{\cos x}{x} - \sin x \right).
\end{equation}
The Bessel functions of the first kind $J_{n+1/2}(x)$ ($n \geq 1$) are determined by the relation
\begin{equation}
J_{n+1/2}(x) = \left[ \frac{2}{\pi x} + Y_{n+1/2}(x) J_{n-1/2}(x)
\right] / Y_{n-1/2}(x),
\label{bess}\end{equation}
where
\begin{equation}\label{j0}
J_{1/2}(x)  = \sqrt{\frac{2}{\pi x}} \sin x.
\end{equation}
We want to remark that equation~(\ref{bess}) gives wrong results if $n \gg x$, but this is not
the case for standard Mie calculations.
In order to avoid an overflow in the forward recursion, a normalization
is used.  The description of the general method of calculations of
Bessel functions can be found in Loskutov~(1971).

The extinction, scattering, backscattering
and radiation pressure efficiency factors
($Q_{\rm ext}$, $Q_{\rm sca}$, $Q_{\rm bk}$ and $Q_{\rm pr}$) are given by
the relations
\begin{equation}
Q_{\rm ext} = \frac{2}{x^2}
\sum_{n=1}^\infty (2n+1) {\rm Re}\{a_n + b_n\},
\end{equation}
\begin{equation}
Q_{\rm sca} = \frac{2}{x^2}
\sum_{n=1}^\infty (2n+1) (|a_n|^2 + |b_n|^2),
\end{equation}
\begin{equation}
Q_{\rm bk} = \frac{1}{x^2}
\left|\sum_{n=1}^\infty (2n+1) (-1)^n (a_n - b_n)\right|^2,
\end{equation}
and
\begin{equation}
\begin{array}{l}
Q_{\rm pr} = Q_{\rm ext} - g Q_{\rm sca}\\ 
\hspace*{7.5mm} = Q_{\rm ext} -  \\
\hspace*{9mm} \displaystyle  \,\,\,\, \frac{4}{x^2}
\left\{\sum_{n=1}^\infty \left[
 \frac{n(n+2)}{n+1} {\rm Re}(a_{n} a_{n+1} + b_{n} b_{n+1}) +
 \frac{2n+1}{n(n+1)} {\rm Re}(a_{n} b_{n}) \right]
\right\}.
\end{array}
\end{equation}
Following quantities can be derived from the above efficiency factors:
\xbeq
\begin{array}{lcp{1cm}l}
{\rm Absorption\mbox{ } efficiency}     & : &  & Q_{\rm abs} = Q_{\rm ext} - Q_{\rm sca},\\
{\rm Albedo}                            & : &  & \Lambda = {Q_{\rm sca}}/{Q_{\rm ext}},\\
{\rm Asymmetry \mbox{ } parameter}      & : &  & g = (Q_{\rm ext} - Q_{\rm pr}) / Q_{\rm sca}.\\
\end{array}
\xeeq
The asymmetry parameter $g$ (or ``mean cosine'' $\langle \cos \Theta \rangle$)
describes the
distribution of the scattered radiation in the forward / backward direction.
It is defined as
\begin{equation}
g = \langle \cos \Theta \rangle = \displaystyle  \frac
{\displaystyle \int_{4 \pi} F_{11}\, \cos \Theta \, {\rm d}\omega}
{\displaystyle \int_{4 \pi} F_{11}\, {\rm d}\omega} =
\frac
{\displaystyle \int_0^{\pi} F_{11}\, \cos \Theta \, \sin \Theta \, {\rm d}\Theta }
{\displaystyle \int_0^{\pi} F_{11}\, \sin \Theta \,{\rm d}\Theta }
\end{equation}
and can be also found from the integration of the element $F_{11}$
of the scattering matrix.

The corresponding cross sections can be derived from the relation
\begin{equation}
C = G Q,
\end{equation}
where $G = \pi a^2$ is the geometrical cross section of the particle.

For the description of scattering by an ensemble of particles consisting
of several species and particles sizes, weighted mean values
of the different quantities described above can be defined.
They are obtained as the sums of the corresponding quantity
averaged over the corresponding size distribution.
Assuming ``J'' different species in the mixture having the fractional
abundances $f_j$ and a particle number density with
some size distribution $n(a)$, we can formulate the following
normalization condition:
\begin{equation}\label{weight}
 \sum^{\rm J}_{j=1} f_j \int^{a_{{\rm max}}}_{a_{{\rm min}}} n(a)\, {\rm d}a = 1,
 \qquad
 \sum^{\rm J}_{j=1} f_j = 1,
\end{equation}
where $a_{{\rm min}}$ and $a_{{\rm max}}$ are the minimum and maximum particle radius.
The Stokes parameters as well as all cross sections are additive.
Therefore, the ensemble averaged values can be derived from their weighted contributions
(see, e.g.,  Martin~1978, \u{S}olc~1980):
\begin{equation}\label{cextmean}
  \langle C_{\rm ext, \, sca, ...} \rangle =
  \sum^{\rm J}_{j=1} f_j \int^{a_{{\rm max}}}_{a_{{\rm min}}}
  n(a) C_{{\rm ext, \, sca, ...,j}}(a)\, {\rm d}a,
\end{equation}
\begin{equation}
  \langle F_{ik} \rangle = \sum^{\rm J}_{j=1} f_j \int^{a_{{\rm max}}}_{a_{{\rm min}}}
  n(a) \hat{F}_{ik, \,j}(a)\, da.
\end{equation}
For the albedo $\Lambda$ and the asymmetry parameter $g$
the following expressions must be used:
\begin{equation}\label{albmean}
  \langle \Lambda \rangle =
\frac{\displaystyle \sum^{\rm J}_{j=1} f_j \int^{a_{{\rm max}}}_{a_{{\rm min}}}
    n(a) C_{{\rm sca},\,j}(a) \, da}
  {\displaystyle \sum^{\rm J}_{j=1} f_j \int^{a_{{\rm max}}}_{a_{{\rm min}}}
    n(a) C_{{\rm ext},\,j}(a)\, da}
  =\frac{\langle C_{\rm sca} \rangle}{\langle C_{\rm ext} \rangle},
\end{equation}
\begin{equation}\label{gmean}
\displaystyle   \langle g \rangle =
\frac{\displaystyle \sum^{\rm J}_{j=1} f_j \int^{a_{{\rm max}}}_{a_{{\rm min}}}
    n(a) C_{{\rm sca},\,j}(a) g_j(a)\, da}
  {\displaystyle \sum^{\rm J}_{j=1} f_j \int^{a_{{\rm max}}}_{a_{{\rm min}}}
    n(a) C_{{\rm sca},\,j}(a)\, da}.
\end{equation}

\subsection{Code description}

The presented Mie scattering code is written in Fortran~90/95.
The main program regulates the input / output of data, calls the routine for
the calculation of Mie scattering characteristics for a single size parameter and particle composition
(embedded in the {\sl module mie\_routines}),
and performs the averaging in case of a polydisperse ensemble.

The calculations are performed with double precision accuracy using the internally defined data type {\tt r2}
\begin{equation}
{\tt integer, parameter, public ::  r2=selected\_real\_kind(9)}
\end{equation}
in the {\sl module~datatype}. Changing the definition of this data type allows one to adapt
the variables to any required number of significant digits.

The routines {\sl shexqnn2} and {\sl aa2} are adaptations of the Mie scattering code
published by Voshchinnikov~(2004).
They were extended in order to calculate the
amplitude functions (see Eqs.~(\ref{s1}), (\ref{s2}))
and to satisfy Fortran~90/95 standards.
The numerical realization for calculation of the scattering amplitude
functions
follows the standard approach (Eqs.~(\ref{pi})--(\ref{pi1})).
We also successfully tested the implementation of a direct calculation
of the Legendre functions, based on the routine {\tt mlpmn.for} 
provided by Zhang \& Jin~(1996), but decided not to include it in
the presented version of the code because it resulted in longer runtimes. 
The runtime of the code is inversely
proportional to the step of the scattering angles $\Theta$
for which the  amplitude functions
and, based on them, the scattering matrix elements are calculated.
The angular step size $\Delta \Theta$ is an input parameter.

The calculation of Mie series for a single particle radius
for a given material is finished
as soon as the relative contribution of the current term to the extinction efficiency 
becomes smaller than $10^{-15}$. If the default maximum number of terms ($2\times10^7$)
is too small to achieve this accuracy, it may be increased in {\tt subroutine shexqnn2}:

{\tt
\hspace*{1cm}! Maximum number of terms to be considered\\
\hspace*{1cm}nterms = 20\,000\,000
}

In order to derive the optical properties of the particle ensemble
weighted over a size distribution,
an arbitrary number of particle radii to be considered can be defined.
The considered radii are equidistantly distributed on a logarithmic scale
within the radius interval $[a_{{\rm min}}$, $a_{{\rm max}}]$.
In the published version, the size distribution follows a power-law
\begin{equation}\label{sd_pwl}
n(a) \propto a^q,
\end{equation}
where $n(a)$ is the relative number of particles with the radius $a$ and
$q$ is a constant, usually negative quantity for a given component of the mixture.
This size distribution was introduced by Mathis et al.~(1977)
for silicate--graphite mixture with
$q=-3.5$ in the process of the interpretation of the interstellar extinction curve and has been frequently
used in different astrophysical applications.
The code may be easily adapted to other particle size distributions by modification
of the following program line (representing the Eq.~(\ref{weight}) combined with
Eq.~(\ref{sd_pwl}))
\begin{equation}
{\tt weight\ =\ abun(jcomp)\ *\ rad**q\ *\ delrad},
\end{equation}
where {\tt weight} is the weight for the component \#{\tt jcomp} with the radius {\tt rad}
and relative abundance {\tt abun}. 
The quantity {\tt delrad} is the radius step width at the current particle radius {\tt rad}
during the numerical integration stated in Eq.~(\ref{weight}).

An arbitrary number of chemically different components in the ensemble can be considered.
Beside the optical properties, each  component is characterized by its
relative abundance in the particle ensemble.
The complex refractive index as a function of wavelength has to be provided
in tabular form for each  component in a separate file in the directory
{\tt ./ri-data/} (see Tabl.~\ref{directory}):

\begin{tabular}{lcll}
 Column 1 & : & Wavelength $\lambda$ $[\mu{\rm m}]$,&   \\
 Column 2 & : & ${\rm Re}\{m(\lambda)\}$,           &   \\
 Column 3 & : & ${\rm Im}\{m(\lambda)\}$            &   \\
\end{tabular}\\

The results are stored in the directory {\tt ./results/} 
(see Tabl.~\ref{directory}, \ref{output}).

The order of the scattering matrix elements in the respective files (see Table~\ref{output})
is defined by the following algorithm:

{\tt
\begin{tabbing}
 \quad \=\quad \=\quad \=\quad \=\quad \=\quad \=\quad \kill
 do {\sl for all wavelengths $\lambda\ [\mu$m$]$, in increasing order}\\
    \> write $\lambda$ \\
    \> do {\sl for all angles $\Theta$; in increasing order}\\
    \>    \> write $\Theta\ [^{\rm o}]$, $F_{ik}(\Theta,\lambda)$\\
    \> end do\\
 end do\\
\end{tabbing}
}

\begin{table}
  \caption{Directory Structure.}
  \begin{tabular*}{\textwidth}{l@{\extracolsep{\fill}}l}
    \toprule
    Directory & Contents\\
    \midrule
    ./        & Executable ({\bf \em miex}), Source code, Makefile\\
    ./ri-data & Tables with optical data\\
    ./results & Output files (see Tab.~\ref{output})\\
    \bottomrule    
  \end{tabular*}
  \label{directory}
  \bigskip
  \bigskip
\end{table}

\begin{table}
\caption{Output files ({\tt project} is the project name).}
\bigskip
  \begin{tabular*}{\textwidth}{l@{\extracolsep{\fill}}l}
    \toprule
Filename & Contents\\
\midrule
{\sl project}             &  Main file containing all results\\
                          &  (see Appendix~\ref{testrun} for an example)\\
{\sl project}{\tt .qext } &  $\lambda$ $[\mu$m$]$, $Q_{\rm ext}$\\
{\sl project}{\tt .cext } &  $\lambda$ $[\mu$m$]$, $C_{\rm ext}$ $[{\rm m}^2]$\\
{\sl project}{\tt .qsca } &  $\lambda$ $[\mu$m$]$, $Q_{\rm sca}$\\
{\sl project}{\tt .csca } &  $\lambda$ $[\mu$m$]$, $C_{\rm sca}$ $[{\rm m}^2]$\\
{\sl project}{\tt .qbk  } &  $\lambda$ $[\mu$m$]$, $Q_{\rm bk}$\\
{\sl project}{\tt .cbk  } &  $\lambda$ $[\mu$m$]$, $C_{\rm bk}$ $[{\rm m}^2]$\\
{\sl project}{\tt .qabs } &  $\lambda$ $[\mu$m$]$, $Q_{\rm abs}$\\
{\sl project}{\tt .cabs } &  $\lambda$ $[\mu$m$]$, $C_{\rm abs}$ $[{\rm m}^2]$\\
{\sl project}{\tt .alb  } &  $\lambda$ $[\mu$m$]$, albedo\\
{\sl project}{\tt .g    } &  $\lambda$ $[\mu$m$]$, asymmetry parameter $g$\\
{\sl project}{\tt .qpr  } &  $\lambda$ $[\mu$m$]$, $Q_{\rm pr}$\\
{\sl project}{\tt .f11  } &  $\Theta\ [^{\rm o}]$, $F_{11}(\lambda\ [\mu$m$],\Theta\ [^{\rm o}])$\\
{\sl project}{\tt .f12  } &  $\Theta\ [^{\rm o}]$, $F_{12}(\lambda\ [\mu$m$],\Theta\ [^{\rm o}])$\\
{\sl project}{\tt .f33  } &  $\Theta\ [^{\rm o}]$, $F_{33}(\lambda\ [\mu$m$],\Theta\ [^{\rm o}])$\\
{\sl project}{\tt .f34  } &  $\Theta\ [^{\rm o}]$, $F_{34}(\lambda\ [\mu$m$],\Theta\ [^{\rm o}])$\\
\bottomrule
\end{tabular*}
\label{output}
\bigskip
\bigskip
\end{table}

\clearpage

{\bf Acknowledgement}\\ \\
This work was supported through the German Research Foundation 
(Emmy Noether Research Program WO 857/2-1 / Research unit 388, Laboratory Astrophysics),
the NASA grant NAG5-11645, 
and the SIRTF legacy science program through an award issued by JPL/CIT under NASA contract 1407.

\clearpage

\begin{appendix}

\newpage
\section{Test Run}\label{testrun}

In the following, {\bf\em miex} is used to calculate scattering matrix elements,
efficiency factors, cross sections, the albedo and the asymmetry parameter
for a particle ensemble typical for astrophysical applications (see Table~\ref{ex-1a} and \ref{ex-1b}):
\begin{enumerate}
\item Three components:
      \begin{enumerate}
      	\item Astronomical silicate
		(relative abundance: 62.5\%; \\ 		~~~input file: {\tt silicate})
	\item Graphite; refractive index for the electric field vector parallel
		to the crystallographic c axis 
		(relative abundance: 25\%; input file: \\{\tt grap-per})
	\item Graphite; refractive index for the electric field vector perpendicular
		to the crystallographic c axis 
		(relative abundance: 12.5\%, input file: {\tt grap-par})
      \end{enumerate}
\item Grain size distribution:
	\begin{itemize}
	\item minimum/maximum grain size:
		$a_{\rm min}$=0.005\,$\mu$m, $a_{\rm max}$=100\,$\mu$m
	\item $n(a) \propto a^{-3.5}$
	\item number of considered, logarithmically equidistantly distributed
	 	radii: 100
	\end{itemize}
\item Calculation of the scattering matrix elements for the angles
	0$^{\rm o}$, 2$^{\rm o}$, 4$^{\rm o}$, \ldots, 180$^{\rm o}$
\end{enumerate}

Input files (with the dust properties) must be located in the directory ``{\tt ./input/}'',
while the results ared stored in the directory ``{\tt ./results/}''.

In a second example, documented in Table~\ref{ex-2a} and \ref{ex-2b}, a dust grain ensemble
with the chemical components as before but with a single grain size of 10\,cm is considered.

The calculations are carried out for 100 wavelengths distributed equidistantly
on a logarithmic scale within the interval [0.05$\mu$m,\,2000$\mu$m].
The optical constants have been interpolated at the particular wavelengths using
the data from Weingartner \& Draine~(2001) and Laor \& Draine~(1993).

We want to remark, that a large variety of  refractive indexes  can be found in the
{\em Jena -- Petersburg Database of Optical Constants} (Henning et al.~1999, see also
Voshchinnikov~2004). The provided dust parameter files are presented in the format
required by {\bf\em miex}.

{\tt \small
\begin{table}
\caption{Input dialog for ``example1''. The runtime of the code on an Intel XEON CPU 3.0\,GHz,
using the Intel Fortran Compiler 8.0, amounts to 19 seconds (required memory: 1.5\,MByte).}
\bigskip
  \begin{tabular*}{\textwidth}{lcl@{\extracolsep{\fill}}}
    \toprule
Real refractive index of the surrounding medium             & : & {\bf 1.0}\\
Number of wavelengths                                       & : & {\bf 100}\\
Number of          components                               & : & {\bf 3}\\
Name of the dust data files (lambda/n/k data)               &   & \\
\hspace*{1cm}001. component : {\bf silicate}                &   & \\
\hspace*{1cm}002. component : {\bf grap-per}                &   & \\
\hspace*{1cm}003. component : {\bf grap-par}                &   & \\
Relative abundances of the different components [\%]        &   & \\
\hspace*{1cm}001. component : {\bf 62.5}                    &   & \\
\hspace*{1cm}002. component : {\bf 25}                      &   & \\
\hspace*{1cm}003. component : {\bf 12.5}                    &   & \\
-1- Single grain size                                       &   & \\
-2- Grain size distribution                                 & : & {\bf 2}\\
\hspace*{1cm}Minimum grain size [micron]                    & : & {\bf 0.005}\\
\hspace*{1cm}Maximum grain size [micron]                    & : & {\bf 100.0}\\
\hspace*{1cm}Size distribution exponent                     & : & {\bf -3.5}\\
\hspace*{1cm}Number of size bins                            & : & {\bf 100}\\
Calculate scattering matrix elements (0=n/1=y)              & : & {\bf 1}\\
\hspace*{1cm}Number of scattering angles in the interval    &   & \\
\hspace*{1cm}$[0^{\rm o}$,180$^{\rm o}]$;  odd number!      &   & \\
\hspace*{1cm}$[$example: '181' $\rightarrow$ step width = 1$^{\rm o}]$ & : & {\bf 91}\\
Project name (8 characters)                                 & : & {\bf example1}\\
Save results in separate files (0=n/1=y)                    & : & {\bf 1}\\
\bottomrule
\end{tabular*}\label{ex-1a}
\end{table}
}

{\tt
\begin{table}
\caption{Program output. File: {\tt ``./results/example1''}}
  \begin{tabular*}{\textwidth}{l@{\extracolsep{\fill}}}
    \toprule
 \# *** PROJECT PARAMETERS ***\\
 \# Number of wavelengths            :          100 \\
 \ldots \\
\midrule
 \# *** RESULTS *** \\
 \# 1. Wavelength [micron], Extinction efficiency factor / cross section [m**2] \\
5.000000000000000E-002   1.32305406861068       4.681097343048097E-016\\
\ldots \\
 \# 2. Wavelength [micron], Scattering efficiency factor / cross section [m**2] \\
  5.000000000000000E-002  0.587523440006963       2.078716568917760E-016\\
\ldots \\
 \# 3. Wavelength [micron], Backscattering efficiency factor / cross section [m**2] \\
  5.000000000000000E-002  6.859423237184992E-002  2.426932402252230E-017\\
\ldots \\
 \# 4. Wavelength [micron], Absorption efficiency factor / cross section [m**2] \\
  5.000000000000000E-002  0.735530628603716       2.602380774130337E-016\\
\ldots \\
 \# 5. Wavelength [micron], Albedo \\
  5.000000000000000E-002  0.444066084633951\\
\ldots \\
 \# 6. Wavelength [micron], Scattering asymmetry factor g \\
  5.000000000000000E-002  0.800843488171790    \\
\ldots \\
 \# 7. Radiation pressure efficiency factor Qpr\\
  5.000000000000000E-002  0.852539747532814\\     
\ldots \\
 \# 8. Wavelength [micron], theta [degree], F11-F12-F33-F34 \\
  5.000000000000000E-002  0.000000000000000E+000   105571.611539073\\
  5.000000000000000E-002  0.000000000000000E+000 -5.641242912656195E-012\\
  5.000000000000000E-002  0.000000000000000E+000   105571.611539073\\
  5.000000000000000E-002  0.000000000000000E+000  7.328649202947496E-011\\
\ldots \\
\bottomrule
\end{tabular*}\label{ex-1b}
\end{table}
}


{\tt \small
\begin{table}
\caption{Input dialog for ``example2''. The runtime of the code on an Intel XEON CPU 3.0\,GHz,
using the Intel Fortran Compiler 8.0, amounts to 129 seconds (required memory: $\sim$240\,MByte).}
\bigskip
  \begin{tabular*}{\textwidth}{lcl@{\extracolsep{\fill}}}
    \toprule
Real refractive index of the surrounding medium             & : & {\bf 1.0}\\
Number of wavelengths                                       & : & {\bf 100}\\
Number of          components                               & : & {\bf 3}\\
Name of the dust data files (lambda/n/k data)               &   & \\
\hspace*{1cm}001. component : {\bf silicate}                &   & \\
\hspace*{1cm}002. component : {\bf grap-per}                &   & \\
\hspace*{1cm}003. component : {\bf grap-par}                &   & \\
Relative abundances of the different components [\%]        &   & \\
\hspace*{1cm}001. component : {\bf 62.5}                    &   & \\
\hspace*{1cm}002. component : {\bf 25}                      &   & \\
\hspace*{1cm}003. component : {\bf 12.5}                    &   & \\
-1- Single grain size                                       &   & \\
-2- Grain size distribution                                 & : & {\bf 1}\\
Grain size [micron]                                         & : & {\bf 100000}\\
Calculate scattering matrix elements (0=n/1=y)              & : & {\bf 0}\\
Project name (8 characters)                                 & : & {\bf example2}\\
Save results in separate files (0=n/1=y)                    & : & {\bf 1}\\
\bottomrule
\end{tabular*}\label{ex-2a}
\end{table}
}

{\tt
\begin{table}
\caption{Program output. File: {\tt ``./results/example2''}}
\bigskip
  \begin{tabular*}{\textwidth}{l@{\extracolsep{\fill}}}
    \toprule
 \# *** PROJECT PARAMETERS ***\\
 \# Number of wavelengths            :          100 \\
 \ldots \\
\midrule
 \# *** RESULTS *** \\
 \# 1. Wavelength [micron], Extinction efficiency factor / cross section [m**2] \\
  5.000000000000000E-002   1.34425083271595       4.223088540642379E-002\\
\ldots \\
 \# 2. Wavelength [micron], Scattering efficiency factor / cross section [m**2] \\
  5.000000000000000E-002  0.731763632480501       2.298903251964923E-002\\
\ldots \\
 \# 3. Wavelength [micron], Backscattering efficiency factor / cross section [m**2] \\
  5.000000000000000E-002  0.174029454773743       5.467296566254283E-003\\
\ldots \\
 \# 4. Wavelength [micron], Absorption efficiency factor / cross section [m**2] \\
  5.000000000000000E-002  0.612487200235445       1.924185288677456E-002\\
\ldots \\
 \# 5. Wavelength [micron], Albedo \\
  5.000000000000000E-002  0.544365392731083   \\  
\ldots \\
 \# 6. Wavelength [micron], Scattering asymmetry factor g \\
  5.000000000000000E-002  0.904868766536477   \\  
\ldots \\
 \# 7. Radiation pressure efficiency factor Qpr\\
  5.000000000000000E-002  0.682100777197063  \\   
\ldots \\
\bottomrule
\end{tabular*}\label{ex-2b}
\end{table}
}

\end{appendix}
\end{document}